\documentclass[fleqn,twoside]{article}
\usepackage{ign2}

\usepackage{graphicx}
\usepackage[figuresright]{rotating}

\newcommand{\AmS}{{\protect\the\textfont2
  A\kern-.1667em\lower.5ex\hbox{M}\kern-.125emS}}
\newcommand{\bbi}{\bibitem}
\newcommand{\ep}{$e^+e^-$}
\title{Standard Model Predictions for the Muon $(g-2)/2$}

\author{S. Eidelman\address[MCSD]{Budker Institute of Nuclear Physics, \\ 
        11 Lavrentyev St., \\ 
        630090 Novosibirsk, Russia}\thanks{Talk at the X Workshop
on $\tau$ Lepton Physics, Novosibirsk, September 2008}}%
       
\begin{document}

\begin{abstract}

The current status of the Standard Model predictions for the muon
anomalous magnetic moment is described.
Various contributions 
expected in the Standard Model are discussed.  After the reevaluation
of the leading-order hadronic term based on the new \ep data,
the theoretical  prediction is more than three standard deviations
lower than the experimental value.

\vspace{1pc}
\end{abstract}

\maketitle

\section{Introduction}
For a particle with spin $\vec{s}$ and magnetic moment $\vec{\mu}$
\begin{equation}
\vec{\mu} =  g \frac {e} {2m} \vec{s},
\end{equation}
where $e,~m$ and $g$ are the charge, mass and gyromagnetic factor
of the particle. In the Dirac theory of a charged pointlike spin-1/2 
particle, $g=2$, and QED effects slightly increase the $g$ value.
Conventionally, a  quantity  $a \equiv (g - 2)/2$ is referred 
to as the anomalous magnetic moment. \\
\indent
The electron and muon anomalous magnetic moments have been measured
with a very high relative accuracy of 0.24 ppb~\cite{gabr} and 
0.54 ppm~\cite{bnl}, respectively. The theoretical prediction
for $a_e$ is only mildly affected by strong and weak interactions providing
a test of QED and giving the most precise value of the fine-structure 
constant $\alpha$. In contrast, $a_{\mu}$ allows to test
all sectors of the Standard Model since all of them contribute
significantly to the total. \\  
\indent
Although the electron anomalous magnetic moments is known much 
more precisely, $a_{\mu}$  is much more sensitive to new physics 
effects: the gain is usually 
$\sim (m_{\mu}/m_{\rm e})^2 \approx 4.3 \cdot 10^{4}$.
The $\tau$ lepton magnetic anomaly has  even better potential, but 
because of the small lifetime of the $\tau$,
it has not yet been measured with the best limits coming from 
DELPHI~\cite{delphi}: $-0.052 < a_{\tau} < 0.013$ at 95\% confidence 
level. The sensitivity of the DELPHI measurement is still one order of
magnitude worse than the predicted value of $a_{\tau}$~\cite{ep07}. \\
\indent
Any significant difference of 
${a}^{\rm exp}_{\mu}$ from ${a}^{\rm th}_{\mu}$ indicates new physics
beyond the Standard Model (SM). It is conventional to write $a_{\mu}$ as
\begin{equation}
{a}^{\rm SM}_{\mu} ={a}^{\rm QED}_{\mu} +
{a}^{\rm EW}_{\mu} + {a}^{\rm had}_{\mu},
\end{equation}
where the terms correspond to the contributions of Quantum Electrodynamics
(QED), electroweak (EW) and strong (hadronic) interactions. While 
discussing these terms and their precision, it is worth comparing them to
the experimental result~\cite{bnl}: 
\begin{equation}
{a}^{\rm exp}_{\mu} = (11659208.0 \pm 6.3) \cdot 10^{-10}.
\end{equation}
\indent
The QED part is dominated by the lowest-order term, represented by one
graph, first-order in $\alpha$~\cite{schwinger}. The number of diagrams
for the second- and third-order terms is more than 100, but they
(up to $\alpha^3$) are known analytically~\cite{remid,la93,lam96}. 
Taking into account a recent more accurate numerical calculation of  
the  $\alpha^4$ terms~\cite{kn04} and the leading log  $\alpha^5$ 
terms\cite{kn06,kataev,chet,kn08} one obtains
\begin{equation}
{a}^{\rm QED}_{\mu}~=~(116584718.09 \pm 0.14 \pm 0.04) \cdot
  10^{-11},
\end{equation}
where the errors are due to the uncertainties of the ${\cal O}(\alpha^5)$
term and $\alpha$, respectively, and the value of 
$\alpha^{-1}=137.035999084(51)$ from the latest measurement of $a_e$
has been used~\cite{gabr,pas}.
It is worth noting that the
4-loop term equals $38.1 \cdot 10^{-10}$ and is thus six times larger 
than the experimental uncertainty. Therefore, it is clear that its 
calculation as well as that of the 5-loop one is necessary. \\
\indent
The electroweak term is known rather accurately~\cite{ckm96,cmv02}:
\begin{equation}
{a}^{\rm EW}_{\mu}~=~(15.4 \pm 0.1 \pm 0.2) \cdot 10^{-10},
\end{equation}
where the first uncertainty is due to hadronic loops while  
the second one is
caused by the errors of $M_H, M_t$ and 3-loop effects. \\
\indent
The hadronic contribution can also be written as a sum:
\begin{equation}
{a}^{\rm had}_{\mu} ={a}^{\rm had,LO}_{\mu} +
{a}^{\rm had,HO}_{\mu} + {a}^{\rm had,LBL}_{\mu}.
\end{equation}
The dominant contribution comes from the leading-order term,
which using dispersion relations can be written as~\cite{bm61,gr69}
\begin{equation}
{a}^{\rm had,LO}_{\mu}=
\left( \frac{\alpha m_\mu}{3\pi} \right)^2
\int_{4m^2_\pi}^{\infty} ds\: \frac{R(s)\:\hat{K}(s)}{s^2}, 
\end{equation}
where
\begin{equation}
R(s)= \frac {\sigma(e^+e^- \to \rm {hadrons})} 
{\sigma(e^+e^- \to \mu^+\mu^-)},
\end{equation}
and the kernel $\hat{K}(s)$ grows from
0.63 at $s=4m^2_\pi$ to 1 
at $s \to \infty$, $1/{s}^2$ emphasizing the role  
of low energies. Particularly important is 
the reaction $e^+e^- \to \pi^+\pi^-$ with a large 
cross section below 1 GeV. Numerically,
${a}^{\rm had,LO}_{\mu} \approx 700 \times 10^{-10}$~\cite{ej95}, 
so we should know it to at least 1\% to match the experimental
accuracy.

\section{Evaluation of the hadronic term}
\begin{figure}[htb]
\begin{center}
\includegraphics[width=0.45\textwidth]{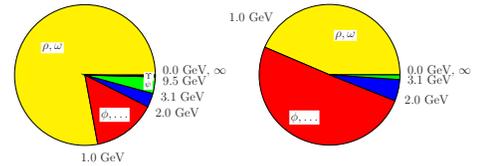}
\caption{Various contributions to  $a^{\rm had,LO}_{\mu}$.}
\label{fig:pie}
\end{center}
\end{figure}

\begin{figure}[htb]
\begin{center}
\includegraphics[width=0.45\textwidth]{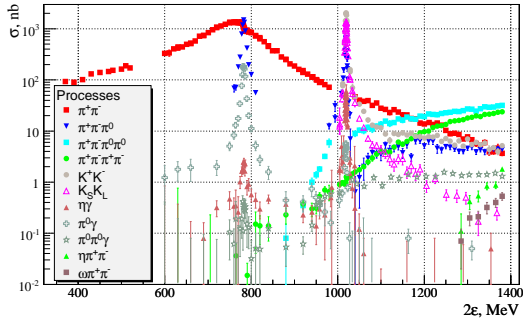}
\caption{Cross sections of various final states from CMD-2.}
\label{fig:all}
\end{center}
\end{figure}

Several estimates of ${a}^{\rm had,LO}_{\mu}$ appeared 
recently~\cite{eid06,hmnt07,j08} based on the progress 
in the low energy \ep annihilation and including the data
not yet available previously~\cite{ej95,dehz1,dehz2}. \\ 
\indent
As already mentioned, one of the largest contributions to 
${a}^{\rm had,LO}_{\mu}$ comes from the $2\pi$ final state 
(about 73\%). Therefore, a high-precision measurement of the
corresponding cross section is one of the main goals of low energy
experiments.   In addition to the previously published $\rho$ meson 
data\cite{cmd04},
CMD-2 reported their final results on the pion form factor $F_{\pi}$ 
from 370 to 1380~MeV\cite{fed,sib,log}. The new $\rho$ meson sample has 
an order of magnitude larger statistics and a systematic 
error of 0.8\%. SND measured $F_{\pi}$ from 390 to 970~MeV with a 
systematic error of 1.3\%\cite{ach}.   
KLOE studied $F_{\pi}$ using the method of radiative return or 
ISR\cite{kho,kuh,ben99} at $590 < \sqrt{s} < 970$~MeV with a sample of
$1.5 \cdot 10^6$~events and systematic error of 1.3\%\cite{kloe}.
The $|F_{\pi}|$ values from CMD-2 and SND are in good agreement. The
KLOE data are consistent with them near the $\rho$ meson peak, but
exhibit a somewhat different energy dependence: they are higher to the
left and lower to the right of the $\rho$ meson peak. However, the 
contributions to $a_{\mu}$ from all three experiments are consistent.
After further analysis of the data with a higher statistics (more than
three million events selected),
KLOE reports the $|F_{\pi}|$ values closer to those of CMD-2 and SND
and achieves a 0.9\% systematic error~\cite{kloe1}. First results 
on the $2\pi$ contribution coming from BaBar will provide additional
information on this channel~\cite{dav08}. \\
\indent
Also important is the contribution of hadronic continuum. Some idea
about the contributions to  $a^{\rm had,LO}_{\mu}$ 
coming from different energy ranges and their uncertainties is given in 
Fig.~\ref{fig:pie}.
Below 1.4~GeV practically all final states have been measured with
consistent results by the CMD-2 and SND groups in Novosibirsk~\cite{igna}.
Figure~\ref{fig:all} shows cross sections of various final states
measured at CMD-2. 
Impressive results on various final states with more than two 
hadrons above 1~GeV were achieved by 
BaBar~\cite{babar1,babar2,babar3,babar4,babar5} using the ISR method. 
In addition to
measuring for the first time the cross sections of new channels, they
show that some of the previous results should be reconsidered. For
example, the BaBar results on the process 
$e^+e^- \to \pi^+\pi^-\pi^0$~\cite{babar1} agree well with those 
of SND~\cite{snd3} below 1.4~GeV, but are considerably higher and
just inconsistent with
those of DM2~\cite{dm2} at higher energies. The overall contribution
of the $\pi^+\pi^-\pi^0$ final state from 1 to 2~GeV (excluding the
$\phi \to \pi^+\pi^-\pi^0$ one) has been 
$(2.45 \pm 0.26 \pm 0.03) \cdot 10^{-10}$
before the BaBar data appeared and it becomes
$(3.25 \pm 0.09 \pm 0.01) \cdot 10^{-10}$ if we replace the DM2 piece
with a much more precise one from BaBar. \\ 
\indent
Using the new data below 1.8~GeV discussed above in addition to the 
whole data set of~\cite{ej95,dehz1} for old experiments, one can reevaluate the 
leading-order hadronic contribution to $a_{\mu}$. The data-based 
calculations~\cite{eid06,hmnt07,j08} slightly differ by the
integration method and the cut-off energy above which the predictions
of perturbative QCD are used, but otherwise are essentially very similar.  
 In Table~\ref{table:t1} we show the results for different
energy ranges following~\cite{eid06}.
 
\begin{table*}[htb]


\hspace{112pt}Table 1

\hspace{112pt}Updated $a^{\rm had,LO}_{\mu}$
\begin{center}
\label{table:t1}     
\newcommand{\m}{\hphantom{$-$}}
\newcommand{\cc}[1]{\multicolumn{1}{c}{#1}}
\renewcommand{\tabcolsep}{2pc} 
\renewcommand{\arraystretch}{1.2} 

\begin{tabular}{cc}
\hline
 $\sqrt{s}$, GeV & \ $a_{\mu}^{\rm had,LO}, 10^{-10}$ \\
\hline
 $2\pi$  & $504.6 \pm 3.1 \pm 1.0$ \\  


 $\omega$ & $38.0 \pm 1.0 \pm 0.3$ \\ 

 $\phi$ & $35.7 \pm 0.8 \pm 0.2$  \\

  $0.6-1.8$ & $54.2 \pm 1.9 \pm 0.4$ \\




 $1.8-5.0$ & $41.1 \pm 0.6 \pm 0.0$   \\

 $J/\psi,\psi'$ & $7.4 \pm 0.4 \pm 0.0$  \\

$> 5.0$ & $9.9 \pm 0.2 \pm 0.0$  \\
\hline
 Total &  $690.9 \pm 3.9_{\rm exp} \pm 2.0_{\rm th}$ \\ 
\hline
\end{tabular}
\end{center}
\end{table*}

\begin{figure}[htb]
\begin{center}
\includegraphics[width=0.45\textwidth]{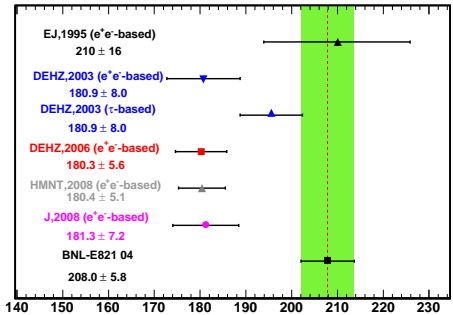}
\caption{Comparison of $a_{\mu}$ from theory and experiment.}
\label{fig:res}
\end{center}
\end{figure}

\begin{figure}[htb]
\begin{center}
\includegraphics[width=0.45\textwidth]{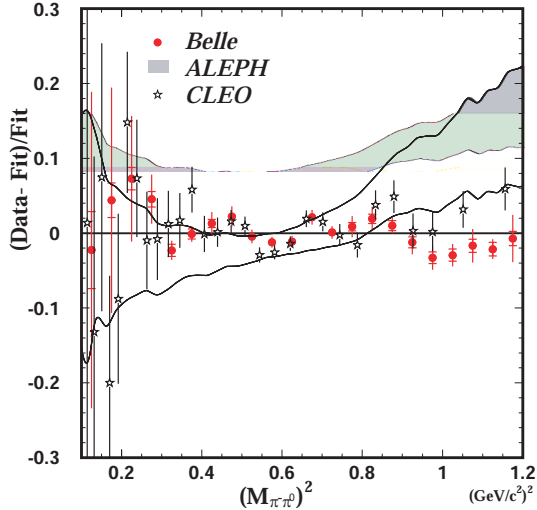}
\caption{Comparison of $2\pi$ spectral functions in $\tau$ decays.}
\label{fig:tau}
\end{center}
\end{figure}

The theoretical error consists of $1.9 \cdot 10^{-10}$ due to 
uncertainties of radiative corrections in old measurements 
and $0.7 \cdot 10^{-10}$
related to using pertubative QCD above 1.8~GeV.
It can be seen that due to a higher accuracy of $e^+e^-$ data the 
uncertainty of $a^{\rm had,LO}_{\mu}$  is now
$4.4 \cdot 10^{-10}$ (0.63\%) compared to $15.3 \cdot 10^{-10}$ 
of Ref.\cite{ej95} and $7.2 \cdot 10^{-10}$ of 
Ref.\cite{dehz2}. \\
\indent
We move now to the higher-order hadronic contributions. Their 
most recent estimate performed in\cite{tt} gives 
${a}^{\rm had,HO}_{\mu}=(-9.8 \pm 0.1) \cdot 10^{-10}$ and  has a 
negligible error compared to that of the leading-order one. \\ 
\indent
The most difficult situation is with the light-by-light hadronic
contribution, which is estimated only theoretically. Even the correct sign 
of this term was established quite recently\cite{kn}.
The older predictions based on the chiral model and vector 
dominance\cite{bpp,hk} were compatible and much lower than that using 
short-distance QCD constraints\cite{mv} (see also\cite{piv}). Their 
approximate averaging in\cite{dm} gives 
${a}^{\rm had,LBL}_{\mu}=(120 \pm 35) \cdot 10^{-11}$. 
Even higher uncertainty is listed in Ref.\cite{bp} who added some terms 
not taken into account in Ref.\cite{mv} to obtain  
$(110 \pm 40) \cdot 10^{-11}$. Two most recent updates give
$(116 \pm 40) \cdot 10^{-11}$~\cite{n09} and  
$(105 \pm 26) \cdot 10^{-11}$~\cite{prv09}.  
It is very tempting
to find an approach to estimate the  light-by-light hadronic
contribution from the data, like, e.g., it was done in Ref.\cite{hkold},
where CLEO single-tag measurements\cite{cleo} of 
$\gamma {\gamma}^* \to \pi^0,\eta,\eta'$ were used to estimate
the contribution from the pseudoscalar resonances. \\
\indent
Using for the light-by-light term the result of Ref.~\cite{dm} and
adding all hadronic contributions, we obtain
$a^{\rm had}_{\mu}=(693.1 \pm 5.6) \cdot 10^{-10}$. 
This result agrees
with other estimations, e.g.,\cite{dehz2,tt,fj,ty,hmnt07,j08}
and its accuracy as well as that of the other recent data-based
evaluations benefits from the new \ep data. \\
\indent
All separate contributions to $a^{\rm th}_{\mu}$ 
are collected in Table~\ref{table:t2}.
\begin{table*}[htb]

\hspace{112pt}Table 2

\hspace{112pt}Experiment vs. Theory
\begin{center}
\label{table:t2}
\newcommand{\m}{\hphantom{$-$}}
\newcommand{\cc}[1]{\multicolumn{1}{c}{#1}}
\renewcommand{\tabcolsep}{2pc} 
\renewcommand{\arraystretch}{1.2} 

\begin{tabular}{cc}
\hline
Contribution &  $a_{\mu}, 10^{-10}$ \\
\hline
Experiment &  $11659208.0 \pm 6.3$ \\
QED &  $11658471.81 \pm 0.02$ \\ 
Electroweak &  $15.4 \pm 0.1 \pm 0.2$ \\  
Hadronic & $693.1 \pm 5.6$ \\
Theory & $11659180.3 \pm 5.6$ \\
\hline
Exp.--Theory &  $27.7 \pm 8.4~(3.3\sigma)$ \\
\hline
\end{tabular}
\end{center}
\end{table*}
Adding the QED, electroweak and hadronic terms, we arrive at the
theoretical prediction of  $(11659180.3 \pm 5.6) \cdot 10^{-10}$. 
The improved precision of the leading-order hadronic contribution
allows to confirm previously observed excess of the experimental value 
of $a_{\mu}$ over the SM prediction with a higher than before 
significance of 3.3
standard deviations. Two other most recent evaluations
also claim a large excess of $3.4\sigma$~\cite{hmnt07} and 
$3.1\sigma$~\cite{j08}. 
Results of the comparison are also shown in Fig.~\ref{fig:res}. 
For the first time during last years the
accuracy of the SM prediction is slightly better than the 
experimental one. \\ 
\indent
How real is a very high accuracy of the leading-order hadronic 
contribution obtained above? We believe that we understand well the 
radiative corrections due to initial-state radiation and vacuum 
polarization, but should not 
forget that they are numerically rather large and may reach $\sim 20\%$,
so their critical reanalysis and tests of the existing Monte 
Carlo generators are needed. The situation with the radiative corrections
due to final state radiation is not so well established, so we 
have to rely on the model of scalar electrodynamics and confront it 
with the data. This may increase the uncertainty of $a^{\rm had,LO}_{\mu}$.
There is
also a question of double counting of the hadronic final states in the
leading- and higher-order hadronic terms\cite{kur}. \\
\indent 
One of the serious experimental questions is that of the missing 
states. An obvious candidate is final states with neutral 
particles only, which have been badly measured before. Recent 
experiments in Novosibirsk
in which the $\pi^0\gamma,~\eta\gamma,\pi^0\pi^0\gamma,~\eta\pi^0\gamma$
final states were studied in the energy range from threshold 
to 1.4~GeV by CMD-2 and SND (see Refs.\cite{rad1,rad2} and 
references therein)
showed that  the cross sections are dominated by the
$\rho, \omega, \phi$~mesons and thus the corresponding contributions 
are properly taken into account. From the upper limits on nonresonant 
cross sections obtained in these papers we can estimate that a possible,
not yet accounted for contribution is  
$a^{\rm had,LO}_{\mu} < 0.7 \cdot 10^{-10}$. However, one should remember 
that there are no measurements at all of the cross sections of such 
channels above 1.4~GeV although they are expected to be small. \\    
\indent
We have already mentioned serious progress with ISR studies from BaBar. 
The discussion of their effect on the $a^{\rm had,LO}_{\mu}$ estimation 
can be subdivided into two parts: new results on already measured states
and studies of various new final states. In the first part there are  
processes which cross sections are consistent with the older 
measurements and more precise, e.g., 
$2\pi^+2\pi^-$,~$\pi^+\pi^-2\pi^0$, $\ldots$. There are also final states
for which the cross sections strongly differ from the older, less accurate
measurements, e.g., $\pi^+\pi^-\pi^0$,~$6\pi$,~$\ldots$. 
In the second part
there are final states, which have never been measured before, e.g.,
$K^+K^-\pi^0\pi^0,~K^+K^-\pi^+\pi^-\pi^0,~4\pi^{\pm}\eta,~K^+K^-\eta$. 
Obviously, one should calculate what contribution to  $a^{\rm had,LO}_{\mu}$
comes from them and add it to the previous estimate. While doing that one
should be very careful since any final state observed may be only
a subset of more general processes. For example, 
the $K^+K^-\pi^+\pi^-\pi^0$ 
final state may come from the process $\phi\eta$, so that our estimate 
of the contribution to $a^{\rm had,LO}_{\mu}$ should be correspondingly 
divided by the relevant branching fractions, in this case 
${\cal B}(\phi \to K^+K^-){\cal B}(\eta \to \pi^+\pi^-\pi^0)=0.1118$,
effectively increasing our estimate of this contribution by a factor 
of 8.94! Fortunately, 
we are interested in exclusive cross sections only below 2~GeV and the
new processes above usually have a rather small cross section in this 
energy range. The first estimate shows that these new contributions 
may increase $a^{\rm had,LO}_{\mu}$ by (1--3)$\cdot 10^{-10}$, 
only slightly decreasing the discrepancy between the theoretical 
expectation and the experimental result. \\
\indent
In view of the new measurements of the cross sections of the processes
with $K^+K^-$ and pions in the final state one should carefully 
reconsider the contribution from the $K\bar{K}n\pi$ final states, 
which was previously 
estimated using isospin relations\cite{dehz1}. Anyway, it is clear that
we have to process new information thoroughly and understand
the size and accuracy of the continuum contribution below 2 GeV 
(now $(62.4 \pm 2.0 \pm 0.5) \cdot 10^{-10}$) compared to
that from the $\pi\pi$ (now $(504.6 \pm 3.1 \pm 1.0) \cdot
10^{-10}$). \\
\indent
There is still no explanation for the observed discrepancy between
the predictions based on $\tau$ lepton and \ep data\cite{dehz2}. For
this reason we are not using $\tau$ data in this update. One expected
that more light
on the problem would be shed by the high-statistics measurement of the
two-pion spectral function by Belle  which preliminary results
indicated to better agreement with \ep data than before\cite{fuji}.
However, it turns out that while in a relatively small range of masses
from 0.8 to 1.2 GeV the $\pi\pi$ spectral function measured at Belle 
is below the ALEPH one, see Fig.~\ref{fig:tau}, this effect is 
compensated by the spectral
function behavior at low and high masses, so that the resultant
contribution to the hadronic part of the muon anomaly is about the
same as before. 
On the other hand, a recent comprehensive analysis of the \ep data
below 1~GeV and those on the $2\pi$ decay of the $\tau$ lepton 
performed in Ref.~\cite{bena} shows that two data sets can be reconciled 
if mixing between the $\rho,~\omega,~\phi$ mesons is taken into account 
in a consistent way. \\   
\indent
Another interesting insight into the problem of the muon anomaly
has been demonstrated in Ref.~\cite{mass}. The authors discuss the possibility
that the observed discrepancy between the experimental measurement of
$a_{\mu}$ and its theoretical prediction may be due to hypothetical errors 
in the determination of the leading-order hadronic contribution. In particular,
they show that if one tries to solve the problem by increasing the low energy
\ep cross section by an amount necessary to bridge the discrepancy, this 
affects the hadronic contribution to the running fine-structure constant 
decreasing the electroweak upper bound on the Higgs mass.
As a result, it leaves a very narrow window for the Higgs mass. They 
also showed that this scenario would require a serious revision of the
\ep cross section, which seems unlikely at the current level of experimental
accuracy. \\    
\indent
What is the future of this SM test? From the experimental side
there are suggestions to improve the accuracy by a factor of 2.5 at
E969 (BNL) or even by an order of magnitude at JPARC. It is clear that
it will be extremely difficult to improve significantly the existing 
accuracy of the leading-order hadronic contribution by measuring the
cross section of \ep annihilation to better than 0.3\% as required
by future determinations of $a_{\mu}$ mentioned above. One can 
optimistically expect substantial progress from
new high-statistics ISR measurements at KLOE, BaBar and Belle together
with the more precise determination of $R$ below 4-5~GeV 
from CLEO-c~\cite{cleorc} and BES-III~\cite{bes}.
Experiments are planned at the new machine 
VEPP-2000 now commissioning, which is a VEPP-2M upgrade up to 
$\sqrt{s}$=2~GeV with $L_{\rm max}=10^{32}$ cm$^{-2}$s$^{-1}$,
with two detectors (CMD-3 and SND)~\cite{vp2000}. 
A similar machine (DA$\Phi$NE-II) is discussed in Frascati~\cite{dafne2}. 
 We can estimate that
by 2010 the accuracy of $a^{\rm had,LO}_{\mu}$ will be improved from
$4.4\cdot 10^{-10}$ by a factor of about 2 (to $\sim 2.2 \cdot 10^{-10}$) 
and the total error of $4.1 \cdot 10^{-10}$ will be limited by the LBL
term ($3.5 \cdot 10^{-10}$), still higher than
the expected $2.5 \cdot 10^{-10}$ in E969. \\ 
\indent
Let us hope that progress of theory will allow a calculation of
$a^{\rm had}_{\mu}$ from first principles (QCD, Lattice). One can
mention here a new approach in the QCD instanton model\cite{dor}
or calculations on the lattice, where there are encouraging
estimates of $a^{\rm had,LO}_{\mu}$, e.g.,\cite{aub} 
$(667 \pm 20) \cdot 10^{-10}$ or attempts 
to estimate $a^{\rm had,LBL}_{\mu}$~\cite{hay}, see also
Ref.~\cite{blum} discussing successes and difficulties of this approach.\\ 
\indent
In conclusion, I'd like to emphasize once again that 
BNL success stimulated significant progress of $e^+e^-$ experiments 
and related theory. Improvement of  $e^+e^-$ data (BaBar, BES, CMD-2, 
KLOE and SND) led to substantial
decrease of the $a^{\rm had,LO}_{\mu}$ uncertainty. For the first time the
accuracy of the theoretical prediction is better than that of the 
experimental measurement. Future experiments as well as development of
theory should clarify whether the observed difference between
$a^{\rm exp}_{\mu}$ and $a^{\rm th}_{\mu}$ is real and what
consequences for the Standard Model and  for possible New
Physics~\cite{newp} it implies.

\section{Acknowledgments} 
I'm indebted to M.~Davier, F.~Jegerlehner, M.~Passera and
G.~Venanzoni for numerous  useful discussions. 
Thanks are also due to my colleagues from VEPP-2M, CMD-2 and SND 
for the long-term collaboration. I appreciate help of V.A.~Cherepanov. \\
\indent
 This work was supported in part by the grants RFBR 06-02-16156, 
RFBR 07-02-00816, RFBR 08-02-13516, RFBR 08-02-91969, 
INTAS/05-1000008-8328, PST.CLG.980342 and DFG GZ RUS 113/769/0-2.

\end{document}